\documentclass[aps,pra,superscriptaddress,12pt]{revtex4-1}

\usepackage{amsmath,amsfonts,amssymb,amsthm}
\usepackage{physics}
\usepackage{mathtools}
\usepackage{setspace}
\usepackage[normalem]{ulem}
\usepackage{stmaryrd}
\usepackage{expdlist}
\usepackage{ytableau}
\usepackage[all]{xy}
\usepackage{nicefrac}

\usepackage{graphicx,xcolor}
\usepackage{hyperref}
\usepackage{url}

\usepackage{verbatim}
\usepackage{alltt}
\usepackage{moreverb}

\usepackage{xr}

\usepackage{enumitem}

\providecommand{\ignore}[1]{}

\newif\ifcmnt
\cmnttrue
\ifdefined\cmntsoff\cmntfalse\fi
\ifcmnt
    \providecommand{\aucmnt}[1]{#1}
    
\else
    \providecommand{\aucmnt}[1]{}
    
\fi

\providecommand{\ignore}[1]{}

\newif\ifcmnt
\cmnttrue
\ifdefined\cmntsoff\cmntfalse\fi

\ifcmnt
    \providecommand{\aucmnt}[1]{#1}

\else
    \providecommand{\aucmnt}[1]{}

\fi

\newcommand{\rls}{\mathbb{R}}
\newcommand{\cmplx}{\mathbb{C}}

\newcommand{\nats}{\mathbb{N}}

\newcommand{\cH}{\mathcal{H}}

\newtheorem{theorem}{Theorem}

\newcommand{\setydiagramtext}{\ytableausetup{nosmalltableaux}\ytableausetup{boxsize=0.4em,centertableaux}}
\newcommand{\setydiagrameq}{\ytableausetup{nosmalltableaux}\ytableausetup{smalltableaux,nocentertableaux}}
\newcommand{\subydiagramtext}[1]{{\ytableausetup{boxsize=0.2em,nocentertableaux}\ydiagram{#1}\setydiagramtext}}
\newcommand{\subydiagrameq}[1]{{\ytableausetup{boxsize=0.2em,nocentertableaux}\ydiagram{#1}\setydiagrameq}}
\newcommand{\parts}{\#}
\newcommand{\diag}{\text{diag}}

\theoremstyle{definition}

\begin{document}

\title{Approximate exchange-only entangling gates for the
three-spin-$1/2$ decoherence-free subsystem}

\author{James R. van Meter}
\affiliation{Department of Mathematics, University of Colorado, Boulder, Colorado 80309, USA}
\affiliation{National Institute of Standards and Technology, Boulder, Colorado 80305, USA}

\author{Emanuel Knill}
\affiliation{National Institute of Standards and Technology, Boulder, Colorado 80305, USA}
\affiliation{Center for Theory of Quantum Matter, University of Colorado, Boulder, Colorado 80309, USA}

\begin{abstract}
  The three-spin-$1/2$ decoherence-free subsystem defines a logical
  qubit protected from collective noise and supports exchange-only
  universal gates. Such logical qubits are well-suited for
  implementation with electrically-defined quantum dots.  Exact
  exchange-only entangling logical gates exist but are challenging to
  construct and understand. We use a decoupling strategy to obtain
  straightforward approximate entangling gates. A benefit of the
  strategy is that if the physical spins are aligned, then it can
  implement evolution under entangling Hamiltonians. Hamiltonians
  expressible as linear combinations of logical Pauli products
  not involving $\sigma_{y}$ can be implemented directly.  
  Self-inverse gates that are constructible from these Hamiltonians,
    such as the CNOT, can be implemented without the assumption on the
    physical spins.  We compare the control complexity of
  implementing CNOT to previous methods and
  find that the complexity for fault-tolerant fidelities is
  competitive.
\end{abstract}

\maketitle

\setydiagramtext

\section{Introduction}

Three physical spin-$1/2$ systems support a collective decoherence-free
two-level subsystem (the 3DFS) in the four-dimensional subspace with
total spin $1/2$, where collective noise is induced by global
fields acting with total spin operators along any axis.  
  This motivates the use of the 3DFS as a logical
qubit~\cite{zanardi:qc1999c,knill:qc1999b}.  The 3DFS's observables
are spanned by the three Hermitian and unitary operators exchanging
two of the physical spins, enabling implementation of logical
one-qubit gates.  These operators are referred to as swaps when used
as unitary gates, and as exchange interactions when used as generators
of evolution. This definition of the exchange interaction agrees
  with the conventional one up to a multiple of the identity.  We also
  identify the swap operators with transpositions
  in the representation of the symmetric group that acts by permuting the
  physical spins.  
The best-known application of
the 3DFS is to quantum computing with quantum dots where evolution
under the exchange interaction can be more accessible than universal
one-quantum-dot evolutions~\cite{divincenzo:qc2000b}. 
Indeed the requisite control over three quantum dots has been
proven feasible by 
many experimental demonstrations \cite{PhysRevB.82.075403,Gaudreau,Hsieh_2012,Medford,Enge1500214,Russ_2017,2018arXiv181202693A}.
The application
to quantum dots is further supported by the ability to implement entangling
logical gates between two blocks of physical spins carrying 3DFSs with
exchange interactions only~\cite{divincenzo:qc2000b}, but these gates
require many steps and are often difficult to construct and understand 
from first principles (with a possible exception being \cite{PhysRevA.93.010303}).

In theory, every logical two-qubit gate is generated by exchange
Hamiltonians on the physical spins comprising two 3DFS blocks
\cite{2001PhRvA..63d2307K}.  In practice constructing such gates is
challenging, and constructing them to be independent of the total spin
of the two blocks even more so.  Since each block has spin 1/2, the
total spin of the system is either 0 or 1.  While the action on each
logical qubit of exchange interactions local to each block is
independent of the total spin, the action 
of cross-block exchange interactions is in general spin-dependent.
  Thus constructing unitaries from cross-block exchange
interactions that act on logical qubits independently of the total
spin is nontrivial, but essential for entangling gates that are robust
under collective noise on the total system.

Much of the work on exchange-only gates has focused on the CNOT gate.
The first exchange-only CNOT gate was found numerically by DiVincenzo et al 
\cite{divincenzo:qc2000b}.  
It required a total spin of 1 and was only approximate, albeit very accurate.  
Subsequently, exact CNOT gates from the same sequence of interactions were found in 
\cite{Hsieh:2003:EUG:974713.974717} and \cite{Kawano2005}. 
A simpler, exact spin-dependent CNOT was later found in \cite{shi}.
A spin-independent, exchange-only CNOT gate
was first published by Fong and Wandzura \cite{DBLP:journals/qic/FongW11} (although a solution was also claimed, 
but not given, in \cite{divincenzo:qc2000b}).  
Surprisingly this exact 
spin-independent solution was simpler, in terms of the number of exchange gates and their analytic coefficients, than that of the original spin-dependent 
solutions.  Since then several other spin-independent, exchange-only CNOT gates 
have been found \cite{setiawan,zeuch}.

The question then arises of how to compare the efficiencies of these
various implementations. 
They each consist of a product of gates, each of which exponentiates
a Hamiltonian that depends linearly on exchange interactions.
  One measure of
complexity, then, is to sum the absolute values of the
coefficients on the exchange interactions. 
Taking the absolute value assumes that the exchange
  interaction can be turned on with either sign
(which is permissible~\cite{PhysRevB.59.2070,BAGRAEV20031061,2008PhRvB..77e4517H,2009PhRvB..80s5103O,2014EL....10547006F,fbf90b1cbccd41febafd7ae35a2b6e87}).  If the
exchange interactions are all performed in series, then this measure
is proportional to the total evolution time
required for the physical operation.  Alternatively, if some exchange
interactions are performed in parallel, one could instead sum the
largest coefficients for each exponential in the
  product for the gate, and again this sum is
proportional to the total evolution time when operating in parallel
mode.  By either measure, the Fong and Wandzura 
construction~\cite{DBLP:journals/qic/FongW11}
  is one of the most efficient of the exact, exchange-only
CNOT gates, and we use it as a benchmark. 

We give an alternative approach that constructs approximate entangling
logical gates between two 3DFSs.
  The strategy is to decouple cross-block
exchange interactions by within-block operations that project the
cross-block interactions into the computational two-qubit subsystem
determined by the two 3DFSs.  With the orthogonal complement of the
computational subsystem thus decoupled, we proceed to construct 
two-qubit gates from exchange interactions
and further show that a large subset of these gates are spin-independent.
This task is facilitated by allowing non-commuting exchange
interactions to be performed in parallel, as made possible for example by semiconductor
quantum dot technology \cite{PhysRevB.93.121410}.  We then find that
for the spin-independent CNOT gate, an entanglement fidelity better
than $0.99$ can be achieved with a total evolution time that is
shorter than that of the exact
implementations, and slightly shorter
still if spin-dependence is allowed.

This paper is organized as follows.  In Section~\ref{sec:II} we review
the symmetric group and unitary group representation theory needed to
derive our results. In particular, we give explicit maps of the
computational subsystems into the 
  symmetric group representations that
appear.  In Section~\ref{sec:III} we describe our procedure for
decoupling the computational subsystem for two 3DFSs from its
orthogonal, ``leakage'' space with exchange interactions.  In
Section~\ref{sec:IV} we apply this method to construct a
spin-independent exchange-only CNOT gate, and spin-dependent
exchange-only evolutions optimized for total spin $1$.  We conclude
with a discussion and open problems in Section~\ref{sec:V}.

\section{Representation Theory}
\label{sec:II}

Here we briefly review the relevant representation theory; 
  for more comprehensive discussion of symmetric group representations 
see \cite{fulton1991representation},
of unitary group representations see \cite{louck2008unitary},
and of both applied to quantum information see \cite{2005quant.ph.12255H}. 
  The special unitary group of $d\times d$ matrices is denoted
by $\mathrm{SU}(d)$, 
  and the
symmetric group of permutations of $n$ elements is denoted by $S_n$,
where we are interested in $n=3$ or $n=6$.  Irreducible
representations (irreps) of the groups $\mathrm{SU}(d)$ and $S_n$ are
conveniently labeled by partitions of $n$ in a canonical way that is made
clear below.  For this purpose it is convenient
to introduce the notation $\nu\vdash n$ signifying that $\nu$ is a
partition of $n$. 
We also write $|\nu|$ for the size of $\nu$, so $|\nu|=n$,
and $\#(\nu)$ for the number of parts of $\nu$.
Now we have that 
for each $\nu\vdash n$ there is an irrep of $S_n$, which we denote by $V_\nu$, 
and for each
$\nu\vdash n$ such that $\#(\nu)\leq d$ there is an irrep of $SU(d)$,
which we denote by $U_\nu$. 
  By convention, each partition can be explicitly
denoted by a Young diagram, consisting of a row of cells for each
part, arranged in descending order of length, with $n$ cells total.
So for $n=3$ the partitions are 
\setydiagrameq
\begin{equation}
\ydiagram{3},\ydiagram{2,1},\ydiagram{1,1,1} 
\end{equation}
\setydiagramtext
and therefore the irreps of $S_3$ are 
$V_\subydiagramtext{3}$, $V_\subydiagramtext{2,1}$, and
$V_\subydiagramtext{1,1,1}$ and the irreps of $SU(2)$ are
  $U_\subydiagramtext{3}$ and $U_\subydiagramtext{2,1}$. 

A convenient choice of basis vectors for each irrep of $S_n$ is labeled
by Young tableaux, defined as follows.
A Young tableau is a Young diagram in which every cell is
filled in with a number from 1 to $n$, with no repetitions, such that
the numbers are increasing along every row from left to right and
along every column from top to bottom.  So for the Young diagram
$\ydiagram{3}$ the only Young tableau is 
\setydiagrameq
\begin{equation}
\ytableaushort{1 2 3}, 
\end{equation}
\setydiagramtext
and
for the Young diagram $\ydiagram{2,1}$ the Young tableaux are:
\setydiagrameq
\begin{equation}
\ytableaushort{1 3,2}, \ytableaushort{1 2,3}.
\end{equation}
\setydiagramtext

Similarly, a convenient choice of basis vectors for each irrep of $SU(d)$
is labeled by Weyl tableaux (also known as semistandard tableaux), defined as follows.
A Weyl tableau is a Young diagram of at most $d$ rows 
with every cell filled in with a number from 1 to $d$, with repetitions allowed, such that the numbers 
are nondecreasing along rows from left to right and increasing along columns from top to bottom.
To avoid confusion with the Young tableaux, 
and because we only
consider $d=2$, we substitute 1 and 2 in  Weyl tableaux by an up 
arrow and a down arrow, respectively.
So for the Young diagram $\ydiagram{3}$ the Weyl tableaux are:
\setydiagrameq
\begin{equation}
\ytableaushort{\uparrow \uparrow \uparrow}, \ytableaushort{\uparrow \uparrow \downarrow}, \ytableaushort{\uparrow \downarrow \downarrow}, \ytableaushort{\downarrow \downarrow \downarrow}, 
\end{equation}
\setydiagramtext
corresponding to the four possible spins along the $z$ axis of
the spin $3/2$ subsystem of three physical spins.
For the Young diagram $\ydiagram{2,1}$ the Weyl tableaux are: 
\setydiagrameq
\begin{equation}
\ytableaushort{\uparrow \uparrow,\downarrow}, \ytableaushort{\uparrow \downarrow,\downarrow},
\end{equation}
\setydiagramtext
corresponding to the two states of the spin $1/2$ subsystem of three physical spins. 

To complete our specification of each irrep we need to decide on a convention
for how each group element is to act on 
the basis vectors.
Our choice for the symmetric group is 
Young's orthogonal form \cite{young1977collected}, which is particularly well-suited for quantum mechanical applications 
because it maps permutations to unitary matrices, and therefore
maps transpositions to Hermitian 
matrices.
 It is defined on the Young tableau basis as follows.
Let $T$ be a Young tableau and for $1\leq i,j\leq n$ let
$d_T(i,j)\equiv(c_T(j)-r_T(j))-(c_T(i)-r_T(i))$ where
$r_T(k)$ and $c_T(k)$ respectively denote the row and column containing number $k$.
Then denoting the transposition in $S_n$ 
that exchanges $i$ and $j$ by $(i j)$,
and its matrix representation on $V_\lambda$ by $\rho_\lambda((i j))$, we have
\begin{equation}
\rho_\lambda((i\ i+1))T=\frac{1}{d_T(i,i+1)}T+\sqrt{1-\frac{1}{d_T(i,i+1)^2}}(i\ i+1)T,
\end{equation}
where 
$(i\ i+1)T$ denotes direct action on $T$ via exchange of the positions
of $i$ and $i+1$. 
Note that $(i\ i+1)T$ results in an invalid tableau (only) when 
$i$ and $i+1$ are in the same row or column, but then 
$d_T(i,i+1)=\pm 1$ and thus the coefficient of the invalid term vanishes.
Since transpositions of the form $(i\ i+1)$ generate $S_n$,
the above definition is sufficient to
determine the representation of every permutation.

Now consider the Hilbert space $\cH$ of three physical spin-$1/2$ particles. 
By Schur-Weyl duality \cite{10.2307/j.ctv3hh48t}
there is a vector space isomorphism between $\cH$ and 
$(V_\subydiagrameq{3}\otimes U_\subydiagrameq{3})\oplus(V_\subydiagrameq{2,1}\otimes U_\subydiagrameq{2,1})$ 
that commutes with the action of $S_3\times SU(2)$.  
Such an isomorphism is called an intertwiner
with respect to the group $S_3\times SU(2)$. 
We denote the existence of such an intertwiner as follows:
\setydiagrameq
\begin{equation}
\cH\overset{S_3\times SU(2)}{\cong}(V_\subydiagrameq{3}\otimes U_\subydiagrameq{3})\oplus(V_\subydiagrameq{2,1}\otimes U_\subydiagrameq{2,1}).
\end{equation}
\setydiagramtext
The Hilbert space on the right-hand side can further be expressed in terms of tableau basis vectors:
\setydiagrameq
\begin{eqnarray}
V_\subydiagrameq{3}\otimes U_\subydiagrameq{3} &\cong&\text{span}\left\{\ytableaushort{1 2 3}\otimes\ytableaushort{\uparrow \uparrow \uparrow},\ytableaushort{1 2 3}\otimes\ytableaushort{\uparrow \uparrow \downarrow},\ytableaushort{1 2 3}\otimes\ytableaushort{\uparrow \downarrow \downarrow},\ytableaushort{1 2 3}\otimes\ytableaushort{\downarrow \downarrow \downarrow}\right\},\\
V_\subydiagrameq{2,1}\otimes U_\subydiagrameq{2,1}&\cong&\text{span}\left\{\ytableaushort{1 3,2}\otimes\ytableaushort{\uparrow \uparrow,\downarrow},\ytableaushort{1 2,3}\otimes\ytableaushort{\uparrow \uparrow,\downarrow},\ytableaushort{1 3,2}\otimes\ytableaushort{\uparrow \downarrow,\downarrow},\ytableaushort{1 2,3}\otimes\ytableaushort{\uparrow \downarrow,\downarrow}\right\}.
\end{eqnarray}.
\setydiagramtext

The above basis vectors can be related by linear transformation to the conventional
product states of the three physical spins, specified in terms of spin basis, by demanding that each swap of the qubits
should be consistent with the action of the 
representation of each transposition on the tableau basis. 
Here we identify the action of transpositions on the product basis with that of 
swaps in the obvious way:
\begin{equation}
(i j)\ket{b_1\cdots b_i\cdots b_j\cdots b_n}=\ket{b_1\cdots b_j\cdots b_i\cdots b_n}.
\end{equation}
If we further assume the convention that 
the Weyl tableaux
are eigenvectors
of the $z$-component of the spin operator, then
up to an overall phase factor we find
the following correspondence with the two-qubit 3DFS encoding of \cite{divincenzo:qc2000b}: 
\setydiagrameq
\begin{eqnarray}
\ytableaushort{1 3,2}\otimes\ytableaushort{\uparrow \uparrow,\downarrow} \mapsto &\frac{1}{\sqrt{2}}(|010\rangle-|100\rangle)&\equiv |0_L\rangle|\uparrow\rangle\\
\ytableaushort{1 3,2}\otimes\ytableaushort{\uparrow \downarrow,\downarrow} \mapsto &\frac{1}{\sqrt{2}}(|101\rangle-|011\rangle)&\equiv|0_L\rangle|\downarrow\rangle\\
\ytableaushort{1 2,3}\otimes\ytableaushort{\uparrow \uparrow,\downarrow} \mapsto &\frac{1}{\sqrt{6}}(2|001\rangle-|100\rangle-|010\rangle)&\equiv|1_L\rangle|\uparrow\rangle\\
\ytableaushort{1 2,3}\otimes\ytableaushort{\uparrow \downarrow,\downarrow} \mapsto &\frac{1}{\sqrt{6}}(2|110\rangle-|011\rangle-|101\rangle)&\equiv|1_L\rangle|\downarrow\rangle.
\end{eqnarray}
\setydiagramtext
Therefore 
the Young tableaux can be identified with logical 0 ($0_L$) and logical 1 ($1_L$) as indicated
above.
It is now clear in terms of representation theory that the independence of this qubit from collective noise on the spin is due to
independence from the action of $SU(2)$. 

We now consider entangling
two such logical qubits.
For the goal of constructing 
gates, we need to calculate the action of cross-block exchange interactions on these
qubits.  To that end we make use of the following 
theorem, adapted from \cite{2005quant.ph.12255H}:
\begin{theorem}
Given irreps $V_\lambda$ of $S_{|\lambda|}$ and $V_\mu$ of $S_{|\mu|}$
and irreps $U_\lambda$ and $U_\mu$ of $SU(d)$, 
\begin{equation}
(V_\lambda\otimes U_\lambda)\otimes(V_\mu\otimes U_\mu)\overset{S_{|\lambda|}\times S_{|\mu|}\times SU(d)}{\cong}\bigoplus_{\substack{\nu\vdash|\lambda|+|\mu|\\\parts(\nu)\leq d}}c_{\mu\lambda}^{\nu}V_\lambda\otimes V_\mu\otimes U_\nu,
\end{equation}
where the isomorphism is an intertwiner with respect to the indicated 
group action and $SU(d)$ is assumed to act simultaneously on $U_\lambda$ and $U_\mu$ on the left hand side.  Further,
the right hand side is isomorphic to a subspace of the Schur-Weyl sum,
\begin{equation}
\bigoplus_{\substack{\nu\vdash|\lambda|+|\mu|\\\parts(\nu)\leq d}}c_{\mu\lambda}^{\nu}V_\lambda\otimes V_\mu\otimes U_\nu\overset{S_{|\lambda|}\times S_{|\mu|}\times SU(d)}{\hookrightarrow}\bigoplus_{\substack{\nu\vdash|\lambda|+|\mu|\\\parts(\nu)\leq d}}V_\nu\otimes U_\nu,
\end{equation}
where the inclusion map is a term-by-term intertwiner with respect to the 
indicated group action such that 
each Littlewood-Richardson coefficient $c_{\lambda\mu}^\nu$
gives the multiplicity of the image of $V_\lambda\otimes V_\mu$ on the
left hand side in each corresponding $V_\nu$ on the right hand side.
\end{theorem}

For the case of $\lambda=\mu=\ydiagram{2,1}$ and $d=2$, the Littlewood-Richardson coefficients are such that
\setydiagrameq
\begin{equation}
(V_\subydiagrameq{2,1}\otimes U_\subydiagrameq{2,1})\otimes(V_\subydiagrameq{2,1}\otimes U_\subydiagrameq{2,1}) \cong (V_\subydiagrameq{2,1}\otimes V_\subydiagrameq{2,1}\otimes U_\subydiagrameq{3,3})\oplus(V_\subydiagrameq{2,1}\otimes V_\subydiagrameq{2,1}\otimes U_\subydiagrameq{4,2})
\hookrightarrow (V_\subydiagrameq{3,3}\otimes U_\subydiagrameq{3,3})\oplus(V_\subydiagrameq{4,2}\otimes U_\subydiagrameq{4,2})
\end{equation}
\setydiagramtext
where both maps are intertwiners with respect to $S_3\times S_3\times SU(2)$,
and term-by-term the inclusion map implies the following two intertwiners:
\setydiagrameq
\begin{eqnarray}
\label{eq:intertwiner1}
V_\subydiagrameq{2,1}\otimes V_\subydiagrameq{2,1} &\overset{S_3\times S_3}{\hookrightarrow}& V_\subydiagrameq{3,3}\\
V_\subydiagrameq{2,1}\otimes V_\subydiagrameq{2,1} &\overset{S_3\times S_3}{\hookrightarrow}& V_\subydiagrameq{4,2}.
\label{eq:intertwiner2}
\end{eqnarray}
\setydiagramtext
The left hand side encodes our two logical qubits, while its four-dimensional image in the five-dimensional $V_\subydiagrameq{3,3}$ and nine-dimensional $V_\subydiagrameq{4,2}$ is the computational subspace of each.  That these intertwiners are
with respect to $S_3\times S_3$ implies that the actions of transpositions local
to each block, on the computational subspace, are consistent across irreps.   
These maps also determine the actions of cross-block transpositions on
the computational subspace, albeit in an irrep-dependent manner.
This is one of our primary concerns for the remainder of this paper.

Before proceeding it is instructive to consider the $SU(2)$
irreps in order to form a more complete physical interpretation.
Observe that the single Weyl basis vector of $U_\subydiagramtext{3,3}$ is 
\setydiagrameq
\begin{equation}
\ytableaushort{\uparrow \uparrow \uparrow,\downarrow \downarrow \downarrow},
\label{eq:u33}
\end{equation}
\setydiagramtext
and the Weyl basis vectors of $U_\subydiagramtext{4,2}$ are 
\setydiagrameq
\begin{equation}
\ytableaushort{\uparrow \uparrow \downarrow \downarrow, \downarrow \downarrow},
\ytableaushort{\uparrow \uparrow \downarrow \downarrow, \uparrow \downarrow}, 
\ytableaushort{\uparrow \uparrow \downarrow \downarrow, \uparrow \uparrow}.
\label{eq:u42}
\end{equation}
\setydiagramtext
By our convention, these basis vectors are eigenvectors of the $z$
component of the total spin operator, with eigenvalues 
0~(Eq.~(\ref{eq:u33})), and -1, 0, 1~(Eq.~(\ref{eq:u42}))
respectively.  Therefore diagram $\ydiagram{3,3}$ is associated with states of
total spin 0 and diagram $\ydiagram{4,2}$ is associated with states of total
spin 1.

Returning to symmetric group irreps, and henceforth overloading Young
diagrams to denote symmetric group irreps directly, notice that
irrep $\ydiagram{3,3}$ has one extra dimension and irrep $\ydiagram{4,2}$ has five
extra dimensions outside of the images of
representation $\ydiagram{2,1}\otimes\ydiagram{2,1}$ under the intertwiners of
Eqs.~(\ref{eq:intertwiner1}-\ref{eq:intertwiner2}).  
The extra dimension in
irrep $\ydiagram{3,3}$ and one of the extra dimensions in irrep $\ydiagram{4,2}$
each correspond to the image of representation $\ydiagram{3}\otimes\ydiagram{3}$
under the respective intertwiner implied by Theorem 1 for
$\lambda=\mu=\ydiagram{3}$, which is associated with states for which
each block has total spin $3/2$.  The four remaining dimensions of
irrep $\ydiagram{4,2}$ correspond to images of
representations $\ydiagram{3}\otimes\ydiagram{2,1}$ and
$\ydiagram{2,1}\otimes\ydiagram{3}$, which are associated with
products of logical qubit states with spin 3/2 states.  
The multiplicity of each image is one in every case, as determined by the 
Littlewood-Richardson coefficients, and further clarified by the following
vector space isomorphisms:
\setydiagrameq
\begin{eqnarray}
\label{eq:space0}
\ydiagram{3,3} &\cong& \left(\ydiagram{2,1}\otimes\ydiagram{2,1}\right)\oplus\left(\ydiagram{3}\otimes\ydiagram{3}\right)\\
\ydiagram{4,2} &\cong& \left(\ydiagram{2,1}\otimes\ydiagram{2,1}\right)\oplus\left(\ydiagram{2,1}\otimes\ydiagram{3}\right)\oplus\left(\ydiagram{3}\otimes\ydiagram{2,1}\right)\oplus\left(\ydiagram{3}\otimes\ydiagram{3}\right).
\label{eq:space1}
\end{eqnarray}
\setydiagramtext
These various maps are illustrated in Fig.~\ref{fig:spaces}.

\begin{figure}
\setydiagrameq
\begin{displaymath}
\xymatrix
{
 & &\ydiagram{2,1}\otimes\ydiagram{2,1}\ar[dddll]\ar[dddrr] & &  \\
 &&&&\\
 & &\ydiagram{2,1}\otimes\ydiagram{3}\ar[drr] & &  \\
\ydiagram{3,3} &&&& \ydiagram{4,2}\\
 & &\ydiagram{3}\otimes\ydiagram{2,1}\ar[urr] & &  \\
 &&&&\\
 & &\ydiagram{3}\otimes\ydiagram{3}\ar[uuull]\ar[uuurr] & &
}
\end{displaymath}
\setydiagramtext
\caption{The injective intertwiners with respect to $S_3\times S_3$ of
the 4-dimensional representation $\ydiagram{2,1}\otimes\ydiagram{2,1}$,
the 2-dimensional representation $\ydiagram{2,1}\otimes\ydiagram{3}$,
the 2-dimensional representation $\ydiagram{3}\otimes\ydiagram{2,1}$,
and the 1-dimensional representation $\ydiagram{3}\otimes\ydiagram{3}$
into the 5-dimesional irrep $\ydiagram{3,3}$ and the 9-dimensional irrep $\ydiagram{4,2}$.}
\label{fig:spaces}
\end{figure}

For the purpose of constructing gates we need to specify the intertwiners 
$\ydiagram{2,1}\otimes\ydiagram{2,1}\hookrightarrow\ydiagram{3,3}$ and
$\ydiagram{2,1}\otimes\ydiagram{2,1}\hookrightarrow\ydiagram{4,2}$
explicitly.  It suffices to map the Young tableaux basis 
\setydiagrameq
\begin{equation}
\left\{\ytableaushort{1 3,2}\otimes\ytableaushort{4 6,5},\ytableaushort{1 3,2}\otimes\ytableaushort{4 5,6},\ytableaushort{1 2,3}\otimes\ytableaushort{4 6,5},\ytableaushort{1 2,3}\otimes\ytableaushort{4 5,6}\right\}
\end{equation}
\setydiagramtext
of representation $\ydiagram{2,1}\otimes\ydiagram{2,1}$ into irreps $\ydiagram{3,3}$ and $\ydiagram{4,2}$.  
The images of each vector in this basis are 
 determined up to overall phase
by demanding that the action of $S_{\{123\}}\times S_{\{456\}}$ on it be consistent
with the action of $S_{\{123\}}\times S_{\{456\}}$ on the original basis,
where $S_{\{ijk\}}$ denotes permutations of $\{i,j,k\}$.
We then find that for $\ydiagram{2,1}\otimes\ydiagram{2,1} \hookrightarrow \ydiagram{3,3}$,
\setydiagrameq
\begin{eqnarray}
\ytableaushort{1 3,2}\otimes\ytableaushort{4 6,5} &\mapsto& \frac{1}{2}\ytableaushort{1 3 5,2 4 6}-\frac{\sqrt{3}}{2}\ytableaushort{1 3 4,2 5 6},\label{eq:spin0basis1}\\
\ytableaushort{1 3,2}\otimes\ytableaushort{4 5,6} &\mapsto& -\frac{\sqrt{3}}{2}\ytableaushort{1 3 5,2 4 6}-\frac{1}{2}\ytableaushort{1 3 4,2 5 6},\\
\ytableaushort{1 2,3}\otimes\ytableaushort{4 6,5} &\mapsto& \frac{1}{2}\ytableaushort{1 2 5,3 4 6}-\frac{\sqrt{3}}{2}\ytableaushort{1 2 4,3 5 6},\\
\ytableaushort{1 2,3}\otimes\ytableaushort{4 5,6} &\mapsto& -\frac{\sqrt{3}}{2}\ytableaushort{1 2 5,3 4 6}-\frac{1}{2}\ytableaushort{1 2 4,3 5 6},\label{eq:spin0basis4}
\end{eqnarray}
\setydiagramtext
and for $\ydiagram{2,1}\otimes\ydiagram{2,1} \hookrightarrow \ydiagram{4,2}$,
\setydiagrameq
\begin{eqnarray}
\ytableaushort{1 3,2}\otimes\ytableaushort{4 6,5} &\mapsto& \frac{1}{2}\ytableaushort{1 3 5 6,2 4}-\frac{\sqrt{3}}{2}\ytableaushort{1 3 4 6,2 5},\label{eq:spin1basis1}\\
\ytableaushort{1 3,2}\otimes\ytableaushort{4 5,6} &\mapsto& \frac{\sqrt{3}}{6}\ytableaushort{1 3 5 6,2 4}+\frac{1}{6}\ytableaushort{1 3 4 6,2 5}-\frac{2\sqrt{2}}{3}\ytableaushort{1 3 4 5,2 6},\\
\ytableaushort{1 2,3}\otimes\ytableaushort{4 6,5} &\mapsto& \frac{1}{2}\ytableaushort{1 2 5 6,3 4}-\frac{\sqrt{3}}{2}\ytableaushort{1 2 4 6,3 5},\\
\ytableaushort{1 2,3}\otimes\ytableaushort{4 5,6} &\mapsto& \frac{\sqrt{3}}{6}\ytableaushort{1 2 5 6,3 4}+\frac{1}{6}\ytableaushort{1 2 4 6,3 5}-\frac{2\sqrt{2}}{3}\ytableaushort{1 2 4 5,3 6}.\label{eq:spin1basis4}
\end{eqnarray}
\setydiagramtext
The resulting right hand sides give the computational basis in each irrep.
Note the above numbering is consistent with the 
physical qubit ``nearest neighbors" and corresponding product
state ordering assumed by
DiVincenzo et al. \cite{divincenzo:qc2000b,Hsieh:2003:EUG:974713.974717}.

\section{Decoupling}
\label{sec:III}
Each computational basis defined above spans what we call the computational
subspace of each irrep, allowing us to further define the computational
submatrix of each matrix representation of a permutation
as its projection onto the computational
subspace.
We find that these computational submatrices of
representations of transpositions in irreps $\ydiagram{4,2}$ and
$\ydiagram{3,3}$
are rich and varied enough that their linear
combinations yield most of the computational Pauli basis.  
  Therefore if we were able to evolve according to
projections onto the computational subspace using only exchange
interactions, we would have a simple way to construct a wide variety
of Hamiltonians and gates.  
In this section we present just such a method, to good approximation.  
Our strategy 
is to remove transitions between the computational subspace and its complement with the general decoupling procedure introduced in \cite{1999PhRvL..82.2417V,1999PhRvL..83.4888V}.
Decoupling involves interspersing pulses with evolution so that the
resulting effective Hamiltonian preserves the computational subspace while
acting as the projection of the original Hamiltonian on this subspace.

For the purpose
of evolving under operators projected onto the computational subspace,
the symmetric sums of local transpositions prove to be useful:
\begin{eqnarray}
\Sigma_a &\equiv& \frac{1}{3}((12)+(13)+(23))\\
\Sigma_b &\equiv& \frac{1}{3}((45)+(46)+(56)).
\end{eqnarray}
It can be easily verified that $\Sigma_a$ and $\Sigma_b$ each act as zero
on the computational subspace, by accordingly exchanging the
physical spins of the product states that constitute the computational basis.

For the sake of consistency with our representation theory approach we 
introduce the group algebra $\cmplx S_6$, which we have just used implicitly.
This consists of all finite linear combinations of $S_6$,
\begin{equation}
\cmplx S_6 \equiv \{a_1\sigma_1+\cdots+a_m\sigma_m|a_i\in\cmplx,\sigma_i\in S_6,m\in\nats\},
\end{equation}
as for example $\Sigma_a$ and $\Sigma_b$.
The group algebra comes equipped with multiplication, addition, and 
scalar multiplication in the obvious ways.
It is then straightforward to linearly extend Young's orthogonal representation from $S_6$ to $\cmplx S_6$, which we assume henceforth. As expected we find 
$\rho_\lambda\left(\Sigma_a\right)$ and
$\rho_\lambda\left(\Sigma_b\right)$ are each null on the computational
subspace.

 The sums $\Sigma_{a}$ and $\Sigma_{b}$ have additional
properties that make them well-suited to the task of performing
effective projection operations.  As they consist of mutually
nonoverlapping transpositions, they commute, and therefore their
representations can be simultaneously diagonalized.  
We then have for irrep $\ydiagram{3,3}$, 
\begin{eqnarray}
\rho_\subydiagramtext{3,3}(\Sigma_a) &=& \diag(0,0,0,0,1),
\label{eq:spin0localsums}\\
\rho_\subydiagramtext{3,3}(\Sigma_b) &=& \diag(0,0,0,0,1),
\end{eqnarray}
and for irrep $\ydiagram{4,2}$,
\begin{eqnarray}
\rho_\subydiagramtext{4,2}(\Sigma_a) &=& \diag(0,0,0,0,1,1,0,0,1),\\
\rho_\subydiagramtext{4,2}(\Sigma_b) &=& \diag(0,0,0,0,0,0,1,1,1).
\label{eq:spin1localsums}
\end{eqnarray}
where here and henceforth the first four basis vectors of irrep $\ydiagram{3,3}$ are given by the right hand sides of Eqs.~(\ref{eq:spin0basis1}-\ref{eq:spin0basis4}), in that order, the first four basis vectors of irrep $\ydiagram{4,2}$ are given by the right hand sides of Eqs.~(\ref{eq:spin1basis1}-\ref{eq:spin1basis4}), in that order (thus corresponding in both cases to the computational
basis $\{\ket{00}_L,\ket{01}_L,\ket{10}_L,\ket{11}_L\}$),
and the remainder of each basis is such that the following properties hold.
The eigenspace of the final 1 
in every case is the image of representation $\ydiagram{3}\otimes\ydiagram{3}$,
the remaining two-dimensional 1-eigenspace in 
$\rho_\subydiagramtext{4,2}(\Sigma_a)$
is the image of  representation $\ydiagram{3}\otimes\ydiagram{2,1}$, and the remaining two-dimensional
1-eigenspace in $\rho_\subydiagramtext{4,2}(\Sigma_b)$ 
is the image of representation $\ydiagram{2,1}\otimes\ydiagram{3}$. 
So $\Sigma_a$ and $\Sigma_b$ can be interpreted as projectors onto states 
that are disallowed by the computational subspace.
Furthermore $\Sigma_a+\Sigma_b\neq 0$ on the complement of the computational subspace
in both irreps. 

To put these sums to use, 
we define the following unitaries,
\begin{eqnarray}
U_a &\equiv& \exp(i\pi\rho_\lambda(\Sigma_a))\\
U_b &\equiv& \exp(i\pi\rho_\lambda(\Sigma_b)),
\end{eqnarray}
where $\lambda$ equals partition $\ydiagram{3,3}$ or $\ydiagram{4,2}$.
For $\lambda=\ydiagram{3,3}$, 
and again in the diagonalizing basis (which includes the computational basis),
\begin{eqnarray}
U_a &\equiv&\diag(1,1,1,1,-1)\\
U_b &\equiv&\diag(1,1,1,1,-1)
\end{eqnarray}
and for $\lambda=\ydiagram{4,2}$,
\begin{eqnarray}
U_a &\equiv&\diag(1,1,1,1,-1,-1,1,1,-1)\\
U_b &\equiv&\diag(1,1,1,1,1,1,-1,-1,-1).
\end{eqnarray}
It follows that, for all $H\in\rho_\lambda(\cmplx S_6)$
and $\lambda=\ydiagram{3,3}$ or $\ydiagram{4,2}$,
\begin{equation}
\frac{1}{4}(H+U_aHU_a^\dagger+U_bHU_b^\dagger+U_bU_aHU_a^\dagger U_b^\dagger)=D(H),
\end{equation}
where $D$ denotes 
a decoupling function which cancels cross terms
with the computational subspace.
Alternatively we can define
\begin{equation}
U\equiv\exp\left(i\frac{\pi}{2}\rho_\lambda(\Sigma_a+\Sigma_b)\right),
\end{equation}
and obtain in both irreps,
\begin{equation}
\frac{1}{4}(H+UHU^\dagger+U^\dagger HU+U^2H(U^2)^\dagger) =D(H),
\end{equation}
with the choice of unitaries depending on which is more convenient for the problem at 
hand.

In terms of projection operators, 
\begin{equation}
D(H)\equiv \left\{\begin{array}{ll} 
\Pi H\Pi + \Pi_{\subydiagrameq{3}\otimes\subydiagrameq{3}}H\Pi_{\subydiagrameq{3}\otimes\subydiagrameq{3}},&  H\in\rho_\subydiagrameq{3,3}(\cmplx S_6),\\
\Pi H\Pi
+\Pi_{\subydiagrameq{2,1}\otimes\subydiagrameq{3}}H\Pi_{\subydiagrameq{2,1}\otimes\subydiagrameq{3}}
+\Pi_{\subydiagrameq{3}\otimes\subydiagrameq{2,1}}H\Pi_{\subydiagrameq{3}\otimes\subydiagrameq{2,1}}
+\Pi_{\subydiagrameq{3}\otimes\subydiagrameq{3}}H\Pi_{\subydiagrameq{3}\otimes\subydiagrameq{3}},& H\in\rho_\subydiagrameq{4,2}(\cmplx S_6),
\end{array}\right.
\end{equation}
where $\Pi$ projects onto the computational subspace while 
$\Pi_{\subydiagrameq{2,1}\otimes\subydiagrameq{3}}$,
$\Pi_{\subydiagrameq{3}\otimes\subydiagrameq{2,1}}$,
and $\Pi_{\subydiagrameq{3}\otimes\subydiagrameq{3}}$
project onto the subspaces comprising the complement of the computational
subspace as indicated in Fig.~1 and Eqs.~(\ref{eq:space0}-\ref{eq:space1}).
Noting that $\rho_\subydiagrameq{3,3}(\cmplx S_6)\cong M_5(\cmplx)$,
if we let $H=(h_{ij})\in\rho_\subydiagrameq{3,3}(\cmplx S_6)$ then we obtain
\begin{equation}
D(H)=\begin{pmatrix}
h_{11} & h_{12} & h_{13} & h_{14} & 0 \\
h_{21} & h_{22} & h_{23} & h_{24} & 0 \\
h_{31} & h_{32} & h_{33} & h_{34} & 0 \\
h_{41} & h_{42} & h_{43} & h_{44} & 0 \\
0 & 0 & 0 & 0 & h_{55}
\end{pmatrix}.
\end{equation}
On the other hand noting that $\rho_\subydiagrameq{4,2}(\cmplx S_6)\cong M_9(\cmplx)$,
if we let $H=(h_{ij})\in\rho_\subydiagrameq{4,2}(\cmplx S_6)$ then we obtain
\begin{equation}
D(H)=\left(\begin{array}{ccccccccc}
h_{11} & h_{12} & h_{13} & h_{14} & 0 & 0 & 0 & 0 & 0 \\
h_{21} & h_{22} & h_{23} & h_{24} & 0 & 0 & 0 & 0 & 0 \\
h_{31} & h_{32} & h_{33} & h_{34} & 0 & 0 & 0 & 0 & 0 \\
h_{41} & h_{42} & h_{43} & h_{44} & 0 & 0 & 0 & 0 & 0 \\
0 & 0 & 0 & 0 & h_{55} & h_{56} & 0 & 0 & 0 \\
0 & 0 & 0 & 0 & h_{65} & h_{66} & 0 & 0 & 0 \\
0 & 0 & 0 & 0 & 0 & 0 & h_{77} & h_{78} & 0 \\
0 & 0 & 0 & 0 & 0 & 0 & h_{87} & h_{88} & 0 \\
0 & 0 & 0 & 0 & 0 & 0 & 0 & 0 & h_{99}
\end{array}\right).
\end{equation}
\setydiagramtext
We now recognize projections onto the four-dimensional computational subspace,
the two-dimensional subspaces associated with $\ydiagram{2,1}\otimes\ydiagram{3}$ and
$\ydiagram{3}\otimes\ydiagram{2,1}$, and the one-dimensional subspace associated with
$\ydiagram{3}\otimes\ydiagram{3}$, as discussed in Section~II.


Our task, then, is to express the above projection procedure in terms
of physically-realizable, exchange-only operations.
Therefore we ``Trotterize", that is approximate
\begin{equation}
\exp(i\alpha D(H))=\exp(i\alpha\left(\frac{1}{4}\sum_{j=1}^4U_jHU_j^\dagger\right))
\end{equation}
 as a product of unitaries in the form of exponentials
depending on linear combinations of exchange interactions,
where $\alpha$ is some angle 
and $\{U_j\}_{j=1}^4$ is one of $\{1,U_a,U_b,U_aU_b\}$
or $\{1,U,U^\dagger,U^2\}$.
Letting $\{A_j\}_{j=1}^k$ be a set of non-commuting operators we can use for example the zeroth order exponential product expansion,
\begin{equation}
\exp\left(\sum_{j=1}^k A_j\right)=\left(\prod_{j=1}^k\exp\left(\frac{1}{n}A_j\right)\right)^n+O\left(\frac{1}{n}\right),
\end{equation}
or the first order Suzuki-Trotter expansion \cite{10.2307/2033649,suzuki1976},
\begin{equation}
\exp\left(\sum_{j=1}^k A_j\right)=\left(\prod_{j=2}^k\exp\left(\frac{1}{2n}A_{k+2-j}\right)
\exp\left(\frac{1}{n}A_1\right)\prod_{j=2}^k\exp\left(\frac{1}{2n}A_j\right)\right)^n+O\left(\frac{1}{n^2}\right),
\end{equation}
or still higher order approximations. 
Therefore, noting that 
\begin{equation}
\exp(i\alpha U_jHU_j^\dagger)=U_j\exp(i\alpha H)U_j^\dagger,
\end{equation}
and using $\{U_j\}_{j=1}^4=\{U^2,U,1,U^\dagger\}$, we have to zeroth order that
\begin{equation}
\exp(i\alpha D(H))=\left(U^2\exp(i\delta tH)(U^2)^\dagger U\exp(i\delta tH)U^\dagger \exp(i\delta tH)U^\dagger\exp(i\delta tH)U\right)^n+O(\delta t)
\end{equation}
and to first order that
\begin{eqnarray}
\exp(i\alpha D(H)) &=& \left(U^\dagger\exp(i\frac{\delta t}{2} H)U\exp(i\frac{\delta t}{2}H) U\exp(i\frac{\delta t}{2}H)U^\dagger U^2\exp(i\delta tH)(U^2)^\dagger\right.\nonumber\\
&&\left.\times U\exp(i\frac{\delta t}{2}H)U^\dagger \exp(i\frac{\delta t}{2}H)U^\dagger\exp(i\frac{\delta t}{2}H)U\right)^n+O(\delta t^2)\\
 &=& U^\dagger\left(\left(\exp\left(i\frac{\delta t}{2}H\right)U\right)^3\exp(i\delta tH)\left(U^\dagger\exp\left(i\frac{\delta t}{2}H\right)\right)^3\right)^nU+O(\delta t^2).\nonumber
\label{eq:trotter}
\end{eqnarray}
where $\delta t=\alpha/4n$.

\section{Hamiltonian and gate construction}
\label{sec:IV}
\renewcommand{\arraystretch}{1.5}

\subsection{One qubit: Spin-independent Hamiltonians and gates}

We now explore which combinations of exchange interactions yield
useful computational submatrices. 
First we review the relationship
between local exchange interactions and Pauli matrices.  The $X$ and
$Z$ Pauli matrices on the first qubit can be constructed from any two
local exchange interactions within the first block as follows:
\begin{eqnarray}
\left(\begin{array}{cc}
 -\frac{1}{\sqrt{3}} & -\frac{2}{\sqrt{3}} \\
 -1 & 0
\end{array}\right)
\Pi\rho_\lambda\left(\begin{array}{c}
(12)\\(13)
\end{array}\right)\Pi
&=&\left(\begin{array}{cc}
 \frac{1}{\sqrt{3}} & \frac{2}{\sqrt{3}} \\
 -1 & 0
\end{array}\right)
\Pi\rho_\lambda\left(\begin{array}{c}
(12)\\(23)
\end{array}\right)\Pi\nonumber\\
&=&\left(\begin{array}{cc}
 -\frac{1}{\sqrt{3}} & \frac{1}{\sqrt{3}} \\
 1 & 1
\end{array}\right)
\Pi\rho_\lambda\left(\begin{array}{c}
(13)\\(23)
\end{array}\right)\Pi = \left(\begin{array}{c}
 XI\\ ZI
\end{array}\right),
\label{eq:local2pauli}
\end{eqnarray}
where the representation $\rho_\lambda$ and projector $\Pi$ are understood to operate on
each component of each column vector of transpositions.
Thus for example the first row of the left-most term implies
\begin{equation}
-\frac{1}{\sqrt{3}}\Pi\rho_\lambda(12)\Pi-\frac{2}{\sqrt{3}}\Pi\rho_\lambda(13)\Pi=XI
\end{equation}
where the projector appears in this equation only for consistency of
matrix dimensions across the equal sign
(decoupling is not required for single-qubit gates),
and 
by a component of the form $AB$, where $A,B\in\{I,X,Z\}$, we mean a tensor
product of the indicated Pauli matrix on the first qubit with the indicated Pauli matrix on the
second qubit.
Similarly local exchange interactions within the second block yield
Eq.~(\ref{eq:local2pauli}) but with $(12)$, $(13)$, and $(23)$ replaced by
$(45)$, $(46)$, and $(56)$ respectively, and $XI$ and $ZI$ replaced by
$IX$ and $IZ$ respectively.

Now we focus on one of the blocks and so momentarily dispense with tensor products with the identity.  We can obtain any Hamiltonian of the form $aX+bZ$ from local exchange interactions, up to phase. 
A Hamiltonian of the form $aX+bY$ can then be obtained by conjugation
with $\exp(i\frac{\pi}{4}X)$, which transforms $Z$ to $Y$,
and a Hamiltonian of the form $aY+bZ$ can be obtained by conjugation 
with $\exp(i\frac{\pi}{4}Z)$, which transforms $X$ to $Y$.

More generally, for any single-qubit Hamiltonian $H$ and time $t$ it follows
from a Cartan decomposition of $SU(2)$ 
(a generalized Euler angle decomposition of rotations) ~\cite{KHANEJA200111}
that
\begin{equation}
\exp(iHt)=\exp(i\delta I)\exp(i\alpha X)\exp(i\beta Z)\exp(i\gamma X),
\end{equation}
for some $\alpha,\beta,\gamma,\delta\in\rls$, which can then be expressed
in terms of local exchange interactions by Eq.~(\ref{eq:local2pauli}). 
Thus we can construct any single-qubit gate from local exchange interactions. 
Such constructions are independent of the choice
of irrep ($\ydiagram{3,3}$ or $\ydiagram{4,2}$) since the relevant intertwiners
preserve the action of local exchange interactions.

\subsection{Two qubits: Spin-dependent Hamiltonians and spin-independent gates}

Constructing exchange-only Hamiltonians for two-qubits is more challenging.  
From calculations performed in Maple 2016 (see ancillary files \verb|spin0_exchange_gates| and \verb|spin1_exchange_gates|)
we find:
\begin{equation}
\label{eq:swap2pauli}
\left(\begin{array}{ccccccccc}
\frac{1}{5} & \frac{1}{5} & \frac{1}{5} & \frac{1}{5} & \frac{1}{5} & \frac{1}{5} & \frac{1}{5} & \frac{1}{5} & \frac{1}{5} \\
-\sqrt{3} & \sqrt{3} & 0 & -\sqrt{3} & \sqrt{3} & 0 & -\sqrt{3} & \sqrt{3} & 0 \\
-1 & -1 & 2 & -1 & -1 & 2 & -1 & -1 & 2 \\
-\sqrt{3} & -\sqrt{3} & -\sqrt{3} & \sqrt{3} & \sqrt{3} & \sqrt{3} & 0 & 0 & 0 \\
-1 & -1 & -1 & -1 & -1 & -1 & 2 & 2 & 2 \\
\frac{3}{2} & -\frac{3}{2} & 0 & -\frac{3}{2} & \frac{3}{2} & 0 & 0 & 0 & 0 \\
\frac{\sqrt{3}}{2} & \frac{\sqrt{3}}{2} & -\sqrt{3} & -\frac{\sqrt{3}}{2} & -\frac{\sqrt{3}}{2} & \sqrt{3} & 0 & 0 & 0 \\
\frac{\sqrt{3}}{2} & -\frac{\sqrt{3}}{2} & 0 & \frac{\sqrt{3}}{2} & -\frac{\sqrt{3}}{2} & 0 & -\sqrt{3} & \sqrt{3} & 0 \\
\frac{1}{2} & \frac{1}{2} & -1 & \frac{1}{2} & \frac{1}{2} & -1 & -1 & -1 & 2
\end{array}\right)
\Pi\rho_\lambda\left(\begin{array}{c}
(14)\\(15)\\(16)\\(24)\\(25)\\(26)\\(34)\\(35)\\(36)
\end{array}\right)\Pi
=a_\lambda\left(\begin{array}{c}
b_\lambda II\\ IX\\ IZ\\ XI\\ ZI\\ XX\\XZ\\ZX\\ZZ
\end{array}\right),
\end{equation}
where $a_\lambda=b_\lambda=1$ for $\lambda=\ydiagram{4,2}$ and $a_\lambda=-3$ and $b_\lambda=-1/5$ for $\lambda=\ydiagram{3,3}$.
Thus within each irrep we can construct any linear combination of Pauli
Hamiltonians excluding $Y$ by decoupling sums of exchange interactions.
As before, combinations of $X$ and $Y$ or $Z$ and $Y$ on one qubit
can be obtained by appropriate conjugation with
$\exp(i\frac{\pi}{4}(ZI))$,
$\exp(i\frac{\pi}{4}(XI))$, 
$\exp(i\frac{\pi}{4}(IZ))$,
or $\exp(i\frac{\pi}{4}(IX))$. 

For any two-qubit gate $G$ it follows
from a Cartan decomposition of $SU(4)$ that \cite{KHANEJA200111}
\begin{equation}
G=K_1\exp(i\alpha XX+i\beta YY+i\gamma ZZ)K_2,
\end{equation}
for some local unitaries $K_1$ and $K_2$ and real coefficients $\alpha,\beta,\gamma$.
The local unitaries can be expressed in terms of $X$ and $Z$ as discussed
previously.  The remaining exponential can be further decomposed 
since $XX$, $YY$, and $ZZ$ all commute; for
example:
\begin{eqnarray}
G&=&K_1\exp(i\alpha XX)\exp(i\beta YY)\exp(i\gamma ZZ)K_2\\
&=&K_1\exp(i\alpha XX)\exp(i\frac{\pi}{4}(XI+IX))\exp(i\beta ZZ)\nonumber\\
&&\times\exp(-i\frac{\pi}{4}(XI+IX))\exp(i\gamma ZZ)K_2.
\label{eq:gate}
\end{eqnarray}

The above Hamiltonian and gate constructions are irrep-dependent, in general.  
However, consider an exchange-only Hamiltonian $H$ equal to a combination of
the above Pauli terms  in the $\ydiagram{4,2}$ irrep 
that squares to the identity. 
Then using that $H$ equals the same combination of Pauli terms multiplied
by -3 in the $\ydiagram{3,3}$
irrep (neglecting a possible identity term contribution to overall
phase) we have by Euler's formula for matrices that
$\exp(i\frac{\pi}{2} H)$ is irrep-independent.
For example,
\begin{eqnarray}
\exp\left(i\frac{3\pi}{4}\Pi\rho_\subydiagramtext{3,3}((14)-(15)-(24)+(25))\Pi\right) &=& \exp\left(-i\frac{3\pi}{2}XX\right)\\
&=& iXX\nonumber\\
&=& \exp\left(i\frac{\pi}{2}XX\right)\nonumber\\
&=& \exp\left(i\frac{3\pi}{4}\Pi\rho_\subydiagramtext{4,2}((14)-(15)-(24)+(25))\Pi\right)\nonumber
\label{eq:example}
\end{eqnarray}
Similarly the general gate $G$ in Eq.~(\ref{eq:gate}) is irrep-independent for coefficients $\alpha,\beta,\gamma\in\{-\pi/2,0,\pi/2\}$.
This procedure is used for the CNOT gate in the next section.

\subsection{Spin-independent CNOT}

Observe that 
\begin{equation}
\frac{1}{2}(IX-ZX)=\left(\begin{array}{cccc}
0 & 0 & 0 & 0 \\
0 & 0 & 0 & 0 \\
0 & 0 & 0 & 1 \\
0 & 0 & 1 & 0
\end{array}\right).
\end{equation}
Further,
letting 
\begin{equation}
N\equiv\frac{3\sqrt{3}}{4}((15)-(14)+(25)-(24)),
\end{equation}
we immediately obtain from subtraction of the 2nd and 8th rows of Eq.~(\ref{eq:swap2pauli})
that
\begin{equation}
\Pi\rho_\subydiagramtext{4,2}(N)\Pi=\frac{1}{2}(IX-ZX),
\end{equation}
and
\begin{equation}
\Pi\rho_\subydiagramtext{3,3}(N)\Pi=-3\Pi\rho_\subydiagramtext{4,2}(N)\Pi.
\end{equation}
Clearly then
\begin{equation}
\exp\left(i\frac{\pi}{2}\Pi\rho_\subydiagramtext{4,2}(N)\Pi\right)=\exp\left(i\frac{\pi}{2}\Pi\rho_\subydiagramtext{3,3}(N)\Pi\right)
\end{equation}
by the same logic as in the example above (Eq.~(\ref{eq:example})),
albeit with the Hamiltonian $\Pi\rho_\subydiagramtext{4,2}(N)\Pi$ squaring to a 
two-dimensional rather than four-dimensional identity.
Meanwhile $I-(12)$, being local, acts the same on the computational subspace of 
both irreps:
\begin{equation}
\Pi\frac{1}{2}\rho_\subydiagramtext{4,2}(I-(12))\Pi=\Pi\frac{1}{2}\rho_\subydiagramtext{3,3}(I-(12))\Pi=\diag(1,1,0,0).
\end{equation}
Thus by Eq.~(\ref{eq:trotter}) the following expression maps to an approximate CNOT on the computational subspace of both irreps,
\begin{eqnarray}
\label{eq:trotterizedcnot}
&&\exp\left(-i\frac{\pi}{4}(I+(12))\right)U^\dagger\left(\left(\exp\left(i\frac{\delta t}{2}N\right)U\right)^3\exp(i\delta tN)\left(U^\dagger\exp\left(i\frac{\delta t}{2}N\right)\right)^3\right)^nU\nonumber\\
&&\hspace*{1in}={\tt CNOT}+O(\delta t^2),
\end{eqnarray}
where $\delta t=\pi/8n$, and for ease of notation we have dropped the explicit 
$\rho_\lambda$ mapping as well as the decoupled orthogonal projection.

\begin{figure}
\begin{displaymath}
\xymatrix
{
& 1\ar@{-}[dddl]\ar@{.}[d]\ar@{-}[dddr] & \\
& 3\ar@{.}[d] & \\
& 2\ar@{-}[dl]\ar@{-}[dr] & \\
4\ar@{.}[rr]\ar@{.}[dr] && 5\ar@{.}[dl] \\
& 6 &
}
\end{displaymath}
\caption{Qubits arranged so that exchange interactions in the spin-independent CNOT are
between neighbors.  Solid lines denote exchange interactions in the Hamiltonian yielding 
the CNOT submatrix while dotted lines denote local interactions used only for the decoupling procedure.}
\label{fig:plane}
\end{figure}
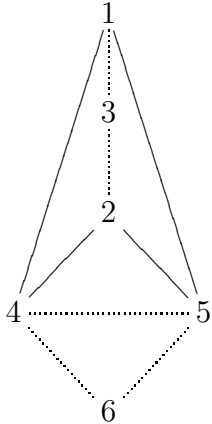

Conveniently, the above Hamiltonian and its Trotterization 
can be recast as neighboring exchange interactions in planar geometry (Fig.~\ref{fig:plane}), where $(12)$ has been removed from $U$ because it commutes
with $N$.  (In general if any one local exchange interaction commutes with the Hamiltonian,
then since it also commutes with the sum of the remaining local interactions it cancels during
conjugation and can thus be removed.)
That $(12)$ commutes with $N$ is due to the fact that
$N$ is symmetric with respect to qubits 1 and 2 and thus invariant
under conjugation with $(12)$.
We add that the exact diagram in Fig.~\ref{fig:plane} may not be physically practical due to the relative 
distances between qubits 1, 4 and 5, 
but more practical architectures may be possible.

To evaluate the efficiency of our approximate CNOT we can 
define a normalized operation time
by summing the maximum magnitude of the coefficients on the exchange interactions 
in each exponential and dividing by $\pi/2$ (the
time required for one full swap).  This normalized time is proportional to the
actual physical time provided the exchange interactions are performed in parallel.
Alternatively we can consider the number of clock cycles, as defined in \cite{divincenzo:qc2000b}, which here equates to the number of exponential factors in Eq.~(\ref{eq:trotterizedcnot}) (after consolidating commuting exponents).
To assess the quality of our approximate CNOT we can use the 
entanglement fidelity \cite{Nielsen:1996uh},
\begin{equation}
F(G,{\tt CNOT})=\left(\frac{1}{4}\text{tr}(\Pi G^\dagger{\tt CNOT})\right)^2.
\label{eq:fidelity}
\end{equation}
We also calculate
the contribution from leakage out of the computational subspace to $1-F$:
\begin{equation}
L(G,{\tt CNOT})=\frac{1}{4}\text{tr}(\Pi G^{\dagger}{\tt CNOT}\Pi^{\perp}{\tt CNOT}^{\dagger}G),
\label{eq:leakage}
\end{equation}
where $G$ is the gate under consideration and {\tt CNOT} has been
extended by identity to the complement of the computational subspace.
Then for various iterations $n$ of the approximation above we 
obtain the values in Table~\ref{table:cnot}, where the fidelity and leakage values assume a 
total spin of 1, this being the worst case (heuristically because its higher dimensional subsystem has more cross-terms to decouple).
\begin{table}
\begin{center}
\begin{tabular}{|c|c|c|c|c|}
\hline
$n$ & cycles &   time & fidelity & leakage \\ \hline
3 & 39 &  8.5   & 0.99136 & 0.00552 \\ \hline
5 & 63 &  12.5 & 0.99888 & 0.00070 \\ \hline
9 & 111 &  20.5 & 0.99989 & 0.00007 \\
\hline
\end{tabular}
\end{center}
\caption{Clock cycles, normalized time (such that one unit equals the duration of one swap), entanglement fidelity (Eq.~(\ref{eq:fidelity})), and leakage (Eq.~(\ref{eq:leakage})) for various iterations $n$ of the spin-independent CNOT approximation given by Eq.~(\ref{eq:trotterizedcnot}).  The listed values for fidelity and leakage assume a total spin of 1. 
The values for cycles and time may be compared with that of the fiducial exact solution
of \cite{DBLP:journals/qic/FongW11}, with 13 clock cycles and
a normalized time of 12.3.}
\label{table:cnot}
\end{table}
If negative exchange interactions are canceled in the optimal way described in Section~\ref{sec:V},
the normalized time is increased by 1.3 in every case
but the fidelities and leakages are unchanged.
The above times are comparable to that of one of the most efficient exact
solutions \cite{DBLP:journals/qic/FongW11}, which has a normalized time of 12.3 (including the local unitaries required for consistency with our computational basis) or 14.8 if negative exchange interactions are canceled.

\subsection{Spin 1 optimized CNOT}

Alternatively the qubits can be prepared with a total spin of 1 
by the use of magnetic fields \cite{divincenzo:qc2000b},
in which case spin-independence is not a concern, and
the $IX$ term in the above derivation of CNOT can be expressed in terms 
of local interactions instead of cross-block interactions.  In combination with the 
noncommuting cross-block interactions in the $ZX$ term, this replacement 
breaks irrep-independence, since the two kinds of interactions transform differently 
across irreps.  However the resulting approximation is made slightly more 
efficient, as we show below.

Let
\begin{equation}
N_1\equiv \frac{\sqrt{3}}{4}((56)-(46)+3((34)-(35))).
\end{equation}
Then for spin 1 we have 
\begin{equation}
\Pi\rho_\subydiagramtext{4,2}(N_1)\Pi={\tt CNOT}.
\end{equation}
Unlike the previous spin-independent expression,
$N_1$ commutes with $U_bN_1U_b^\dagger$ and similarly
$U_aN_1U_a^\dagger$ commutes with $U_aU_bN_1U_b^\dagger U_a^\dagger$.
This simplifies the Suzuki-Trotter approximation for 
$\exp(i\alpha(N_1+U_aN_1U_a^\dagger+U_bN_1U_b^\dagger+U_bU_aN_1U_a^\dagger U_b^\dagger))$ and we find 
\begin{equation}
\label{eq:trotterizedcnot1}
\exp\left(-i\frac{\pi}{4}(I+(12))\right)
(T^{1/2}U_aT^{1/2}U_a^\dagger U_bTU_aTU_b^\dagger T^{1/2}U_a^\dagger T^{1/2})^n={\tt CNOT}+O(\delta t^2),
\end{equation}
where $T\equiv\exp(i\delta tN_1)$ and $\delta t=\pi/8n$.
Again this can be arranged as neighboring exchange interactions in a plane (Fig.~\ref{fig:spin1plane}), where $(12)$ commutes with $N_1$ and so can be
removed from $U_a$.
\begin{figure}
\begin{displaymath}
\xymatrix
{
& 6\ar@{-}[dl]\ar@{-}[dr] & \\
 4\ar@{-}[dr]\ar@{.}[rr] & & 5\ar@{-}[dl] \\
& 3\ar@{.}[dl]\ar@{.}[dr] & \\
1 &  & 2
}
\end{displaymath}
\caption{Qubits arranged so that exchange interactions in the spin-dependent CNOT are
between neighbors.  Solid lines denote exchange interactions in the Hamiltonian yielding
the CNOT submatrix while dotted lines denote local interactions used only for the decoupling procedure.}
\label{fig:spin1plane}
\end{figure}

The smaller truncation error that results from this expression yields the results shown in Table~\ref{table:cnot1}.
\begin{table}
\begin{center}
\begin{tabular}{|c|c|c|c|c|}
\hline
$n$ & cycles & time & fidelity & leakage \\ \hline
2 & 21 & $9.8$ & 0.99849 & 0.00067 \\ \hline
3 & 31 & $13.8$ & 0.99970 & 0.00014 \\ \hline
4 & 41 & 17.8 & 0.99990 & 0.00004 \\
\hline
\end{tabular}
\end{center}
\caption{Clock cycles, normalized time (such that one unit equals the duration of one swap), entanglement fidelity (Eq.~(\ref{eq:fidelity})), and leakage (Eq.~(\ref{eq:leakage})) for various iterations $n$ of the spin 1 CNOT approximation given by Eq.~(\ref{eq:trotterizedcnot1}).  The above values for cycles and time may be compared with that of the fiducial exact solution
of \cite{DBLP:journals/qic/FongW11}, with 13 clock cycles and
a normalized time of 12.3.}
\label{table:cnot1}
\end{table}
Again if subtracted exchange interactions are canceled the normalized time is increased by 1.3 in every case.

\section{Discussion}
\label{sec:V}

For the case of two logical qubits encoded on a decoherence-free subsystem
of three physical spins each, we have presented an exchange-only method
to decouple the resulting computational subspace from leaked states.
Using Trotterization we showed how this method can be
implemented to any desired fidelity.  We then applied this procedure to
explicitly construct Pauli Hamiltonians from exchange interactions.
Building on this we provided a transparent method to construct 
two-qubit gates from exchange interactions,  
including a large family of spin-independent gates.  
In particular we have shown that an exchange-only
CNOT gate can be implemented with a computational speed comparable to or
better than previous solutions, with good fidelity, provided the exchange interactions 
can be performed in parallel. The fidelities for these 
constructions
exceed $0.99$, which is comparable to fidelities demonstrated for
one logical qubit in three quantum dots~\cite{2018arXiv181202693A}, and
for two-qubit gates with other quantum devices~\cite{2014Natur.508..500B,gaebler:qc2016a,PhysRevLett.117.060504}. 
Fidelities exceeding the conservative requirements for fault-tolerance~\cite{Preskill:1997ds,2010Natur.463..441K,2010Natur.464...45L}
can be achieved with a slow down of about a factor of two.

We note that for quantum dots the negative coefficients on exchange 
interactions in
many of the expressions above would seem to require negative charging energies
\cite{PhysRevA.57.120}.  This is possible by coupling to magnetic fields
or other means
\cite{PhysRevB.59.2070,BAGRAEV20031061,2008PhRvB..77e4517H,2009PhRvB..80s5103O,2014EL....10547006F,fbf90b1cbccd41febafd7ae35a2b6e87},
but if it proves impractical, the negative coefficients can be removed
by various methods.
Of course for an exponential of a single negative 
exchange interaction one may simply add an integer multiple of $2\pi$ to its coefficient.

To address the more general case of multiple negative exchange interactions, we 
observe that the sum of all transpositions in $S_6$ acts as a 
constant on the irreps,
\begin{equation}
\rho_\lambda\left(\sum_{j\neq k=1}^6(jk)\right)=c_\lambda I,
\end{equation}
where $c_\subydiagramtext{3,3}=3$ and $c_\subydiagramtext{4,2}=5$.
Therefore adding the sum of all exchange interactions to 
an exchange-only Hamiltonian, with a coefficient equal to the largest 
magnitude of the negative coefficients in that Hamiltonian, will cancel that 
negative 
coefficient and ensure that all other resulting coefficients are nonnegative.  
In general this results in an overall phase difference between the 
spin 0 and spin 1 cases, due to the different constants effectively being added,
but our encoding is independent of such a phase.  Alternatively for 
negative local exchange interactions we can add an appropriate multiple of the 
sum of all local exchange interactions, for one or both blocks as needed, as 
that sum acts as zero on the computational subspace (Eqs.~(\ref{eq:spin0localsums}-\ref{eq:spin1localsums})).  
It follows that for negative cross-block interactions we can add an appropriate 
multiple of the sum of all cross-block interactions, as that sum acts as 
$3I$ or $5I$ on the computational subspace depending again on whether the total spin is 0 or 1.
To summarize in more physical terms,
the sums discussed above commute with the logical two-qubit observables,
which are in the group algebra generated by $S_3\times S_3$, and can therefore be
added to the Hamiltonian as needed without changing relevant expectation values.
 
We end with some open questions for future work.  Aside from the Pauli
and CNOT gates, it may be of interest to find efficient expressions
for other gates and evaluate
their efficiencies as we did for CNOT.  Regarding CNOT, we sought to
minimize the time required for given fidelities, 
but one might instead seek to minimize 
the 
clock cycles or circuit depth.
Related to
this, it may be of practical interest to approximate our parallel
interaction expressions by strictly serial expressions.  Finally,
our decoupling procedure employed a first order Suzuki-Trotter
approximation, which is optimal for the fidelities considered here,
but higher order approximations might be advantageous 
if higher fidelities are desired.
Although the first order approximation can achieve any fidelity
with sufficiently many iterations, the higher order approximations require
fewer iterations and for large enough fidelity they are more efficient.

\begin{acknowledgments}
  JRvM wishes to thank M. Ly, K. Mayer, and N. Thiem for helpful
  discussions.  This work includes contributions of the National
  Institute of Standards and Technology, which are not subject to
  U.S. copyright. 
The identification of any product or trade names
  is for informational purposes and does not imply
  endorsement or recommendation by the National Institute of Standards and Technology. This work was performed under
  the financial assistance award 70NANB18H006 from U.S. Department of 
  Commerce, National Institute of Standards and Technology.
JRvM also acknowledges the support of the Professional Research Experience
Program at the National Institute of Standards and Technology.
\end{acknowledgments}

\bibliographystyle{unsrt}
\bibliography{gates}

\end{document}